\documentclass[twocolumn,showpacs,superscriptaddress,amsmath,amssymb,aps,prb]{revtex4}
\usepackage{epstopdf}
\usepackage[dvips]{graphicx}
\usepackage{amsmath,bm}
\usepackage[dvips]{color}
\usepackage{graphicx}
\usepackage{epsfig}

\begin{document}
\preprint{APS/123-QED}
\title{Controllable exchange coupling between two singlet-triplet qubits}
\author{Rui Li}
\affiliation{Department of Physics and State Key Laboratory of
Surface Physics, Fudan University, Shanghai 200433, China}

\author{Xuedong Hu}
\affiliation{Department of Physics, University at Buf\mbox{}falo, SUNY, Buf\mbox{}falo, New York 14260-1500, USA}

\author{J. Q. You}
\affiliation{Department of Physics and State Key Laboratory of
Surface Physics, Fudan University, Shanghai 200433, China}
\affiliation{Beijing Computational Science Research Center, Beijing
100084, China}

\begin{abstract}
We study controllable exchange coupling between two singlet-triplet qubits. We start from the original second quantized Hamiltonian of a quadruple quantum dot system, and obtain the effective spin-spin interaction between the two qubits using the projection operator method. Under a strong uniform external magnetic field and an inhomogeneous local micro-magnetic field, the effective interqubit coupling is of the Ising type, and the coupling strength can be expressed in terms of quantum dot parameters. Finally, we discuss how to generate various two-qubit operations using this controllable coupling, such as entanglement generation, and a controlled-NOT gate.
\end{abstract}

\date{\today}
\maketitle

\section{introduction}

A quantum computer is more efficient than a classical computer in solving certain problems such as prime factorization.\cite{Shor} A variety of physical systems have been suggested as qubits, the building block of a quantum computer.\cite{Ladd} Spin qubits have attracted wide attention for more than a decade\cite{Hanson, Laird, Morton, Buluta} because of the well-developed spin resonance techniques\cite{Slichter} for coherent control and the strong exchange coupling between electron spins in semiconductors.\cite{Loss} In recent years, there have been a multitude of experiments verifying spin control,\cite{Johnson,
Petta, Koppens, Tarucha, Foletti, Maune, Sachrajda, Shulman} spin coherence,\cite{Johnson, Petta, Koppens, Bluhm, Maune, Lyon} and spin measurement.\cite{Barthel1, Barthel2}

Singlet-triplet qubit ($ST_0$ qubit) is a logical qubit encoded in the two-spin singlet $\frac{1}{\sqrt{2}}(|\!\!\uparrow \downarrow\rangle - |\!\!\downarrow \uparrow \rangle)$ and unpolarized triplet $\frac{1}{\sqrt{2}}(|\!\!\uparrow \downarrow\rangle + |\!\!\downarrow \uparrow \rangle)$ states in a double quantum dot (DQD).\cite{Levy, Taylor} Since both states have zero magnetic quantum number, this qubit is insensitive to noises in magnetic fields that are uniform or slowly varying in space. More importantly, this qubit can be initiated with high fidelity and manipulated with precision.\cite{Petta, Foletti} Its fast
measurement has been demonstrated as well.\cite{Petta, Barthel1} Coupling of two singlet-triplet qubits can be implemented capacitively\cite{Taylor2} or through exchange.\cite{Levy} Recent theoretical\cite{Taylor2, Stepanenko, Ramon1, Ramon2, ENielsen} and experimental\cite{Shulman} explorations are focused on the
capacitive coupling. However, this approach has a sensitive inter-dependence of the two-qubit coupling strength and the single-qubit dephasing rate,\cite{Shulman,Ramon2} with stronger coupling leading to faster dephasing and vice versa. To overcome this problem, additional assistances such as a long-distance coupler have been explored to provide enhanced capacitive coupling without increased dephasing.\cite{Trifunovic}

The difficulty in achieving strong capacitive coupling without increased decoherence prompts us to revisit the coupling scheme in the original proposal, using the exchange interaction between the two $ST_0$ qubits. Here we address one of the main issues raised in the original proposal about the exchange coupling, i.e., the issue that it potentially leads to leakage out of the $ST_0$ qubit subspace. Specifically, we calculate the effective coupling between two $ST_0$ qubits, and identify the conditions under which the qubit leakage can be minimized. We also construct several useful two-qubit gates based on the exchange interaction.

The paper is organized as follows. In Sec.~\ref{Sec_II}, we derive the exchange coupling between two electron spins from the second quantized Hamiltonian of a DQD system. In Sec.~\ref{Sec_III}, we extract a controllable coupling between two $ST_{0}$ qubits from an exchange-coupled quadruple quantum dot system. In Sec.~\ref{Sec_IV}, we study how to generate entanglement and build controlled-NOT gate using this controllable coupling. At last, we give a summary in Sec.~\ref{Sec_V}.

\section{\label{Sec_II}Singlet-triplet qubit in a double quantum dot}

The two-electron spin Hamiltonian has been studied extensively in the past decade,\cite{Loss,Burkard,Hu,Scarola} and its form in the $ST_0$ subspace has also been explored.\cite{Coish} Here we quickly summarize the results as a starting point for our calculations in the next section and derive an analytical expression for the two-electron exchange coupling that includes all the tunable parameters.

As discussed in Appendix~\ref{Appendix A}, we derive the two-electron effective Hamiltonian from a generalized Hubbard model.\cite{Yang,Yang2,Wang} In the so-called (11) regime, where each quantum dot contains one and only one electron, the effective spin interaction Hamiltonian for two electrons in a DQD takes the simple Heisenberg exchange form,
\begin{equation}
H_{\rm eff}=J\textbf{S}_{1}\cdot\textbf{S}_{2},\label{EQ_exchange}
\end{equation}
with the exchange coupling between the two electrons
\begin{equation}
J= \frac{4(t-J_{t})^{2}}{U-U'-|\Delta\varepsilon|}-2J_{e} \,.
\label{eq:exchange constant}
\end{equation}
Here $t$ is the single-electron tunneling across the DQD, $J_t$ is the single-electron tunneling in the presence of a second electron, $U$ is the on-site (i.e., intradot) Coulomb repulsion, $U'$ is the interdot Coulomb repulsion, $\Delta\varepsilon$ is the single electron ground-orbital energy difference between the two dots (interdot bias), and $J_e$ is the direct exchange interaction of the two electrons across the DQD.

As discussed in Appendix~\ref{Appendix A}, the exchange splitting $J$ in Eq.~(\ref{eq:exchange constant}) is derived from the generalized Hubbard model and is quite complete; it includes the effects of interdot tunneling, interdot bias, and both on-site and off-site Coulomb interactions. The expression for $J$ contains two parts. The first part, $4(t-J_{t})^{2} / (U-U'-|\Delta\varepsilon|)$, is the antiferromagnetic superexchange between the two dots. The second part, $-2J_{e}$, is the ferromagnetic direct exchange between the two electrons from their Coulomb interactions. The value of total exchange $J$ can be either positive or negative, depending on the values of the parameters $t$, $J_{t}$, $\Delta\varepsilon$, $J_{e}$, $U$, and $U'$. This switch in the exchange coupling between anti-ferromagnetic and ferromagnetic interactions has been previously observed in calculations based on various levels of molecular orbital approximations.\cite{Burkard, Hu, Scarola} It is important to note here that the zero point for the interdot bias here differs from the convention in the singlet-triplet qubit community. Here $\Delta \varepsilon = 0$ means that the single-electron ground states are on resonance across the two dots, while in the conventional definition the zero point is where the (11) singlet and the (02) singlet states have the same energy when tunneling is neglected:
$\varepsilon_L + \varepsilon_R + U' = 2\varepsilon_R + U$. Thus the two bias definitions are shifted by $U-U'$.

Including the external magnetic field (applied along the $\hat{z}$ direction) and the Overhauser field from the lattice nuclear spins, the complete two-spin Hamiltonian is
\begin{equation}
H=\gamma_{e}BS^{z}_{1}+\gamma_{e}BS^{z}_{2}+J\textbf{S}_{1}
\cdot\textbf{S}_{2}+\textbf{S}_{1}\cdot\textbf{h}_{1}+\textbf{S}_{2}\cdot\textbf{h}_{2},
\end{equation}
where $\gamma_{e}$ is the electron gyromagnetic ratio, and $\textbf{h}_{1(2)}=\sum_{l}A_{l,1(2)}\textbf{I}_{l,1(2)}$ is the nuclear field operator in each dot, with $A_{l,1(2)}$ being the hyperfine coupling between each electron spin and its nuclear spin bath.

When the applied magnetic field is large, $\gamma_{e}B\gg J \,, |\textbf{h}_{1(2)}|$, the singlet $S$ and the unpolarized triplet $T_{0}$ states are isolated from the polarized triplet states $T_{\pm}$. Under this condition, $S$ and $T_0$ can be used as the two basis states to encode a single logical qubit, i.e., the $ST_0$ qubit. Within the $ST_0$ subspace, the qubit dynamics is governed by the Hamiltonian\cite{Coish}
\begin{equation}
H=\frac{J}{2} \tau^{Z} + \delta\, h^{z} \, \tau^{X},\label{Eq_exchage_Hamiltonian}
\end{equation}
where $\tau^{Z}=|T_{0}\rangle\langle\,T_{0}|-|S\rangle\langle\,S|$ and $\tau^{X}=|S\rangle\langle\,T_{0}|+|T_{0}\rangle\langle\,S|$ are the Pauli operators of the $ST_{0}$ qubit, with $X$, $Y$ and $Z$ representing the three axes of the $ST_0$ Bloch sphere. $\delta\,h^{z}=(h^{z}_{1}-h^{z}_{2})/2$ is the Overhauser field difference (in general it is the total magnetic field difference that could come from both the Overhauser field and an externally applied inhomogeneous field) across the DQD along the $\hat{z}$ direction in real space. The exchange splitting $J$ and the magnetic field difference $\delta\, h^z$ provide universal control over a single $ST_0$ qubit.

\section{\label{Sec_III} Controllable coupling of two singlet-triplet qubits}

\begin{figure}
\includegraphics{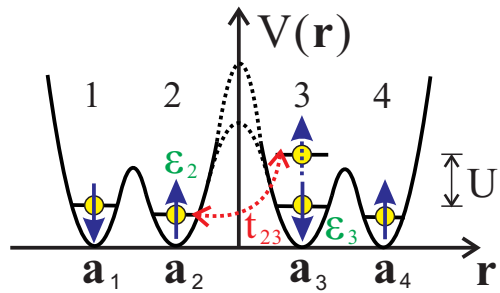}
\caption{\label{Fig_II}(Color online) Four quantum dots are designed
to be coupled as an array, where only single orbital level is
considered in each quantum dot, and on-site (i.e., intradot) Coulomb
repulsion is denoted by $U$. Here we allow the tunneling coupling
$t_{23}$ between dots 2 and 3, which is controlled by the external
gate voltage, i.e., the height of potential between dots 2 and 3.}
\end{figure}

We now study the exchange coupling of two $ST_0$ qubits. The system we consider is a linearly coupled quadruple quantum dot (see Fig.~\ref{Fig_II}), with dots 1 and 2 encoding the first qubit, and dots 3 and 4 encoding the second. In addition to the tunneling between dots 1 and 2, and the tunneling between dots 3 and 4, we further allow tunnel coupling between dots 2 and 3, so the two $ST_0$ qubits are now coupled. This is the main difference between this coupling method and the previous capacitive coupling scheme.\cite{Taylor2, Stepanenko, Ramon1, Ramon2, ENielsen, Shulman}

As discussed in Appendix~\ref{Appendix B}, the effective interaction Hamiltonian for the quadruple quantum dot system is that for a linear Heisenberg spin chain of four nodes:
\begin{equation}
H_{\rm
eff}=J_{12}\textbf{S}_{1}\cdot\textbf{S}_{2}+J_{23}
\textbf{S}_{2}\cdot\textbf{S}_{3}+J_{34}\textbf{S}_{3}
\cdot\textbf{S}_{4},
\label{Hamiltonian-four-spins}
\end{equation}
where the exchange couplings are
\begin{eqnarray}
J_{k,k+1}&=&\frac{4(t_{k,k+1}-J^{k,k+1}_{t})^{2}}{U-U'+U''-|\Delta\varepsilon_{k,k+1}|}-2J^{k,k+1}_{e}.
\label{Eq_exchange_Hamiltonian2-main}
\end{eqnarray}
Here $U$ is the intradot Coulomb repulsion. The interdot Coulomb repulsions are $U_{12}=U_{34}=U'$, and $U_{23}=U''$. The parameters $t_{k,k+1}$, $J^{k,k+1}_{t}$, $J^{k,k+1}_{e}$, and $\Delta\varepsilon_{k,k+1}$ characterize the nearest-neighbour tunnel-coupling, occupation-modulated tunneling, direct exchange, and interdot bias, respectively. The exchange couplings $J_{k,k+1}$ are tunable via $t_{k,k+1}$ and $\Delta\varepsilon_{k,k+1}$, while the Coulomb parameters $U$, $U'$, $U''$, $J_{e}$, and $J_{t}$ cannot be tuned easily (see Figs.~\ref{Fig_II} and \ref{Fig_I}).

After including all the magnetic interaction terms, the total four-spin Hamiltonian reads
\begin{eqnarray}
H_{\rm 6d}&=&\gamma_{e}(B+B_{m})\sum^{2}_{k=1}S^{z}_{k}+\gamma_{e}(B-B_{m})\sum^{4}_{k=3}S^{z}_{k}\nonumber\\
&&+J_{12}\textbf{S}_{1}\cdot\textbf{S}_{2}+J_{23}\textbf{S}_{2}\cdot\textbf{S}_{3}+J_{34}\textbf{S}_{3}\cdot\textbf{S}_{4}\nonumber \\
&&+\sum^{4}_{k=1}\textbf{S}_{k}\cdot\textbf{h}_{k}\,,\label{Eq_coupled_spins1}
\end{eqnarray}
where ${\textbf h}_{k}$ is the nuclear Overhauser field in dot $k$.  We have also introduced a local magnetic field for each $ST_0$ qubit, so spins 1 and 2 are in a local micromagnetic field $\textbf{B}_{m}$, while spins 3 and 4 are in an opposite micromagnetic field $-\textbf{B}_{m}$, in addition to the overall uniform magnetic field $\textbf{B}$.  This local field can be produced by local micromagnets,\cite{Tarucha,Obata,Brunner} and prevents qubit leakage, as discussed below.

In general, the Hilbert space for the four electron spins has a dimension of 16. When a strong uniform external magnetic field $\textbf{B}$ along the $\hat{z}$ direction is applied, i.e., $\gamma_{e}B \gg |{\textbf h}_{k}|$ and $\gamma_{e}B \gg J_{k,k+1}$, spaces with different total magnetic quantum number $S^z = \sum^{4}_{k=1}S^{z}_{k}$ are decoupled.  The $ST_0$ qubits satisfy $S^z = 0$, so we focus here on the $S^z = 0$ sub-Hilbert-space.  It is spanned by $|SS\rangle$, $|ST_{0}\rangle$, $|T_{0}S\rangle$, $|T_{0}T_{0}\rangle$, $|T_{+}T_{-}\rangle$, and $|T_{-}T_{+}\rangle$.  Notice that the two-$ST_{0}$-qubit Hilbert space, spanned by $|SS\rangle$, $|ST_{0}\rangle$, $|T_{0}S\rangle$, $|T_{0}T_{0}\rangle$, has only four dimensions and is only a subspace of the $S^z = 0$ space.  Hamiltonian (\ref{Hamiltonian-four-spins}) shows that states in the $ST_0$ subspace are coupled to $|T_{+}T_{-}\rangle$ and $|T_{-}T_{+}\rangle$ by $J_{23} \textbf{S}_{2}\cdot\textbf{S}_{3}$. To prevent leakage into these two states, we introduce the local magnetic field $\pm \textbf{B}_{m}$ to separate them energetically from the two-qubit Hilbert space. When $J_{23} \ll \, 2\gamma_{e}B_{m} < 2\gamma_{e}B$, the $ST_0$ states would be decoupled from $|T_{+}T_{-}\rangle$ and $|T_{-}T_{+}\rangle$, so two $ST_0$ qubits would evolve without loss when they are coupled through the $J_{23}$-coupling.
%
%This Hamiltonian is defined on the 6 dimensional Hilbert space for the coupled four spins as analyzed above, when the parameters satisfy %$J_{k,k+1}\ll\,2\gamma_{e}B_{m}<2\gamma_{e}B$, the four coupled spins are only allowed to evolve in the product Hilbert space of two $ST_{0}$ qubits (the 4 %dimensional Hilbert space). Therefore,

\begin{figure}
\includegraphics{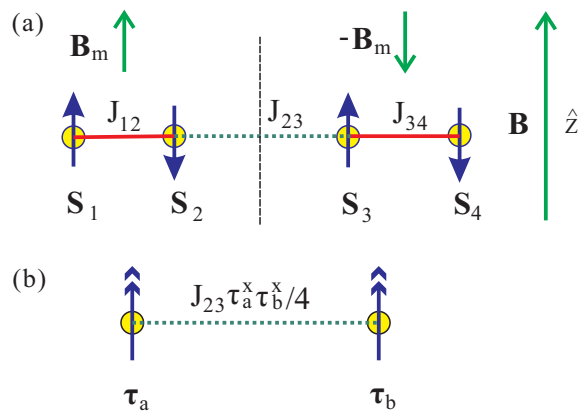}
\caption{\label{Fig_III}(Color online) (a) Exchange-coupled four
electron spins derived from a linearly coupled quadruple quantum dot
system. The exchange couplings $J_{k,k+1}$ are controllable by
external gate voltages. A strong uniform external magnetic field
$\textbf{B}$ is applied along the $\hat{z}$ direction. Also, we
apply local micro-magnetic field to each spin; spins 1 and 2 are in
a local micro-magnetic field $\textbf{B}_{m}$, while spins 3 and 4
are in an opposite micro-magnetic field $-\textbf{B}_{m}$. (b) Under
this carefully designed magnetic field, effectively coupled two
pseudo spins (each pseudo spin represents a $ST_{0}$ qubit) are
extracted from the linearly coupled four spins shown in (a). The
pseudo spin $\boldsymbol{\tau}_{a}$ is formed by $\textbf{S}_{1}$
and $\textbf{S}_{2}$, and $\boldsymbol{\tau}_{b}$ by
$\textbf{S}_{3}$ and $\textbf{S}_{4}$. The coupling between these
two pseudo spins is generated from the exchange coupling $J_{23}$
between spins 2 and 3, and the resulted controllable coupling is of
the Ising XX type.}
\end{figure}

Projecting Hamiltonain (\ref{Eq_coupled_spins1}) into the two-qubit Hilbert space, we obtain the following effective Hamiltonian
\begin{eqnarray}
H_{\rm 4d}&\approx&\gamma_{e}(B+B_{m})\sum^{2}_{k=1}S^{z}_{k}+\gamma_{e}(B-B_{m})\sum^{4}_{k=3}S^{z}_{k}\nonumber\\
&&+J_{12}\textbf{S}_{1}\cdot\textbf{S}_{2}+J_{34}\textbf{S}_{3}\cdot\textbf{S}_{4}+J'_{23}S^{z}_{2}S^{z}_{3}\nonumber \\
&&+\sum^{4}_{k=1}\textbf{S}_{k}\cdot\textbf{h}_{k}\,,\label{Eq_coupled_spins2}
\end{eqnarray}
where $J'_{23}\approx\,J_{23}+(J_{23})^{2}/(4\gamma_{e}B_{m})\approx\,J_{23}$. The qubits are now well defined even when they are coupled, with spins 1 and 2 forming one $ST_{0}$ qubit (denoted as $a$), spins 3 and 4 forming another $ST_{0}$ qubit (denoted as $b$), and the interaction $J'_{23}S^{z}_{2}S^{z}_{3}$ couples these two $ST_{0}$ qubits. Figure~\ref{Fig_III} shows a diagrammatic representation of this projection process. The four linearly-coupled electron spins in the specially designed magnetic field [Fig.~\ref{Fig_III}(a)] are equivalent to two coupled pseudo spins [Fig.~\ref{Fig_III}(b)], with each pseudo spin representing an $ST_{0}$ qubit. Expressing all the electron spin operators of Eq.~(\ref{Eq_coupled_spins2}) in terms of the pseudo-spin operators, we obtain the Hamiltonian of two coupled $ST_{0}$ qubits
\begin{equation}
H=\frac{J_{12}}{2}\tau^{Z}_{a}+\frac{J_{34}}{2}\tau^{Z}_{b}
-\frac{J_{23}}{4}\tau^{X}_{a}\tau^{X}_{b}+\delta\,h^{z}_{a}\tau^{X}_{a}
+\delta\,h^{z}_{b}\tau^{X}_{b}. \label{Eq_coupled_qubtis}
\end{equation}
This is the Hamiltonian for a transverse field Ising model. In other words, tunnel-coupling of two $ST_0$ qubits is equivalent to a controllable effective quantum Ising coupling between them.

For a two-spin qubit, the singlet and triplet states are not the only way to encode a logical qubit. An alternative is to encode the logical qubit in the $|\!\!\uparrow \downarrow \rangle$ and $|\!\!\downarrow \uparrow \rangle$ states. This encoding is less susceptible to charge noise, since the two-electron orbital wave functions and the associated charge distributions are identical for $|\!\!\uparrow \downarrow \rangle$ and $|\!\!\downarrow \uparrow \rangle$ states, but differs for $S$ and $T_0$ states. Indeed, the recent experiment that demonstrated a 200-$\mu$s coherence time in a GaAs double dot is for these two states.\cite{Bluhm} For this encoding scheme, the effective coupling Hamiltonian takes the form
\begin{equation}
H=\frac{J_{12}}{2}\tau^{X}_{a}+\frac{J_{34}}{2}\tau^{X}_{b}
-\frac{J_{23}}{4}\tau^{Z}_{a}\tau^{Z}_{b}+\delta\,h^{z}_{a}\tau^{Z}_{a}
+\delta\,h^{z}_{b}\tau^{Z}_{b}.
\end{equation}
It is still the transverse field Ising model Hamiltonian. Thus the pulse sequences for two-qubit gates would be similar for this encoding as the $ST_0$ encoding.

Notice that the coupling we have here has a form that differs from the capacitively coupled $ST_{0}$ qubits, where the total Hamiltonian corresponds to a longitudinal field Ising model (or classical Ising coupling).\cite{Shulman,Ramon2}

In our exchange coupling scheme, the coupling strength $J_{23}/4$ is only limited by the external fields $J_{23}\ll\,2\gamma_{e}B_{m}<2\gamma_{e}B$, but not by the single-qubit level spacings $J_{12}$ and $J_{34}$. Also, $J_{23}$ can be tuned to be comparable to, or even larger than, $J_{12}$ and $J_{34}$. Indeed, $J_{12}$ and $J_{34}$ are allowed to vanish altogether $J_{12}=J_{34}=0$, and the system will still remain in the $ST_0$ qubit space. This is in strong contrast to the capacitive coupling scheme,\cite{Taylor2,Stepanenko,Ramon1,Ramon2,ENielsen} where the coupling strength is limited by $J_{12}$ and $J_{34}$, with $J_{23}/J_{12}, J_{23}/J_{34}\approx10^{-2}$ in a recent experimental demonstration (see Ref.~\onlinecite{Shulman}).

To estimate the maximum value of the achievable coupling strength of the interqubit coupling $\frac{J_{23}}{4} \tau^{X}_{a}\tau^{X}_{b}$, we take a GaAs quantum dot structure as an example. With current experimental technology, a magnetic field gradient of $\sim20-30$ mT can be realized.\cite{Tarucha,Obata,Brunner} The corresponding Zeeman energy gradient is $2\gamma_{e}B_{m} \approx 120-180$ MHz, so the maximum coupling strength is about $J_{23}/4 \approx 10$ MHz. In principle, the magnetic field gradient $2\gamma_{e}B_{m}$ can be several times larger, so an interqubit coupling close to 100 MHz (0.4 $\mu$eV) should be possible.

\section{\label{Sec_IV}Generating two-qubit gates}

Two-qubit gates are essential for universal quantum computing. In this section, we discuss how to use the Ising interaction derived in the previous section to generate Bell states and to construct a controlled-NOT gate.

\subsection{\label{swap}Entanglement generation via free evolution}

For entanglement generation, we tune the qubit splittings to $J_{12}=J_{34}=J \gg \delta\, h^z$ (putting the two qubits in resonance after neglecting the nuclear fields) and write $J_{23}=J'$. The two-qubit Hamiltonian now takes the simpler form
\begin{equation}
H=\frac{J}{2}\tau^{Z}_{a}+\frac{J}{2}\tau^{Z}_{b}-\frac{J'}{4}\tau^{X}_{a}\tau^{X}_{b}.\label{Eq_coupled}
\end{equation}
The eigenfunctions and the corresponding eigenvalues of this Hamiltonian are
\begin{eqnarray}
|1\rangle\!&\!=\!&\!\cos\frac{\theta}{2}|\!\uparrow_{a}\uparrow_{b}\rangle
+\sin\frac{\theta}{2}|\!\downarrow_{a}\downarrow_{b}\rangle,~~E_{1}=\sqrt{J^{2}+(J'/4)^{2}}, \nonumber\\
|2\rangle\!&\!=\!&\!\sin\frac{\theta}{2}|\!\uparrow_{a}\uparrow_{b}\rangle
-\cos\frac{\theta}{2}|\!\downarrow_{a}\downarrow_{b}\rangle,~~E_{2}=-\sqrt{J^{2}+(J'/4)^{2}}, \nonumber\\
|3\rangle\!&\!=\!&\!\frac{1}{\sqrt{2}}(|\!\uparrow_{a}\downarrow_{b}\rangle
-|\!\downarrow_{a}\uparrow_{b}\rangle),~~E_{3}=\frac{1}{4}J',\nonumber\\
|4\rangle\!&\!=\!&\!\frac{1}{\sqrt{2}}(|\!\uparrow_{a}\downarrow_{b}\rangle
+|\!\downarrow_{a}\uparrow_{b}\rangle),~~E_{4}=-\frac{1}{4}J',
\end{eqnarray}
where $\theta=\arctan(-J'/4J)$. Since $J'$ is controllable, we can turn it on for a period of time $\tau$, and keep it off otherwise. Such a process can be realized in a quantum dot device by tuning the tunneling coupling $t_{23}$ between dots 2 and 3 in Eq.~(\ref{Eq_exchange_Hamiltonian2-main}). If the two qubits are initialized to a desired product state $|\Psi(0)\rangle = |\!\uparrow_{a}\downarrow_{b} \rangle$,\cite{Hanson2,Foletti} after the $\tau$-period evolution, the final state is
\begin{equation}
|\Psi(\tau)\rangle=e^{-iH\tau}|\!\uparrow_{a}\downarrow_{b}\rangle
=\cos\frac{J'\tau}{4}|\!\uparrow_{a}\downarrow_{b}\rangle
+i\sin\frac{J'\tau}{4}|\!\downarrow_{a} \uparrow_{b}\rangle.
\end{equation}
This is generally an entangled state for the two $ST_0$ qubits. In particular, when $J'\tau_{\rm ent}/4=n\pi \pm \pi/4$, two maximally-entangled Bell states are obtained:
\begin{equation}
\frac{\sqrt{2}}{2}(|\!\uparrow_{a}\downarrow_{b}\rangle\pm\,i|\!\downarrow_{a}
\uparrow_{b}\rangle)=e^{-iH\tau_{\rm
ent}}|\!\uparrow_{a}\downarrow_{b}\rangle.
\end{equation}

If the initial state is set to $|\Psi(0)\rangle=|\!\uparrow_{a}\uparrow_{b}\rangle$, free evolution under Hamiltonian (\ref{Eq_coupled}) leads to entangled states between $|\!\uparrow_{a}\uparrow_{b}\rangle$ and $|\!\downarrow_{a}\downarrow_{b}\rangle$. Specifically, when $E_{1}\tau_{\rm ent}=n\pi + \pi/2$, we obtain the following entangled state:
\begin{equation}
\cos\theta|\!\uparrow_{a}\uparrow_{b}\rangle
+\sin\theta|\!\downarrow_{a}\downarrow_{b}\rangle =e^{-iH\tau_{\rm
ent}}|\!\uparrow_{a}\uparrow_{b}\rangle.
\end{equation}
The degree of entanglement in this case depends on the ratio $J'/J$. Since $J'$ can be even larger than $J$, we can imagine tuning of the couplings to $J'=4J$, such that $\theta=-\pi/4$. Now the final state is another maximally-entangled Bell state.

In short, starting from different unentangled initial states, a pulsed $J'$-gate can generate different maximally-entangled Bell states. Since $J'$ is limited by the interdot magnetic field gradient, it may not be feasible to reach the $J' > J$ regime in GaAs while neglecting the local nuclear fields $\delta\, h^z$.  However, this limit should be much easier to achieve in Si, where the nuclear Overhauser field is much smaller.\cite{Assali,Maune}

\subsection{\label{CNOT}Controlled-Not gate}

One of the most commonly used building block for universal quantum circuits is the controlled-NOT (CNOT) gate,\cite{Nielsen} which can be generated relatively straightforwardly using an Ising-type interaction.\cite{Makhlin} In our case, we can adopt two different approaches to construct a CNOT gate. The first approach starts with Eq.~(\ref{Eq_coupled}). After two single-qubit rotations $U_{1}=e^{i\frac{\pi}{4}(\tau^{Z}_{a}+\tau^{Z}_{b})}$ and $U_{2}=e^{-i\frac{\pi}{4}(\tau^{Y}_{a}+\tau^{Y}_{b})}$, Hamiltonian (\ref{Eq_coupled}) can be transformed to $H'$
\begin{equation}
e^{-iH'\tau}=U^{\dagger}_{2}U^{\dagger}_{1}e^{-iH\tau}U_{1}U_{2},
\end{equation}
with
\begin{equation}
H'=-\frac{J}{2}\tau^{X}_{a}-\frac{J}{2}\tau^{X}_{b}-\frac{J'}{4}\tau^{Y}_{a}\tau^{Y}_{b}.
\end{equation}
The pulse sequence of a CNOT gate using this Hamiltonian is well known\cite{Makhlin, Oh}
\begin{equation}
U^{ab}_{\rm CNOT}=H_{b}e^{-i\phi\tau^{Z}_{a}}e^{i\phi\tau^{Z}_{b}}e^{-iH'\tau}e^{-i\pi\tau^{Z}_{a}/2}e^{-iH'\tau}H_{b},
\end{equation}
where $H_{b}=\frac{1}{\sqrt{2}}(\tau^{Z}_{b}+\tau^{X}_{b})$ is the Hadamard gate on qubit $b$, the duration of the two-qubit gate is $\tau=\pi/(4J)\times\sqrt{(4n)^{2}-(2m-1)^{2}}$, where $m$ and $n$ are integers, and $\phi=(\pi/4)(2m-1)$. The coupling parameters must be tuned to the ratio of $J/J'=\sqrt{[4n/(2m-1)]^{2}-1}/4$. Notice that the two-qubit gate is employed twice in this pulse sequence.

Alternatively, we can tune the parameter $J$ in Eq.~(\ref{Eq_coupled}) to $J=0$, which makes the construction of a CNOT gate rather simpler. The local nuclear Overhauser field $\delta\, h^{z}$ can be completely eliminated by a Hahn spin echo.\cite{Hahn} The spin echo (the pulses are simultaneously applied to the two $ST_0$ qubits respectively) under Hamiltonian (\ref{Eq_coupled_qubtis}) leads to a two-qubit quantum gate $e^{i\frac{\pi}{4}\tau^{X}_{a}\tau^{X}_{b}}$, where we have chosen $J'\tau/4=\pi/4$. The CNOT gate can then be realized as follows
\begin{equation}
U^{ab}_{\rm CNOT}=e^{i\frac{\pi}{4}\tau^{Y}_{a}}e^{i\frac{\pi}{4}\tau^{X}_{a}
\tau^{X}_{b}}e^{-i\frac{\pi}{4}(\tau^{X}_{a}+\tau^{X}_{b}-1)}e^{-i\frac{\pi}{4}\tau^{Y}_{a}}.
\end{equation}
Note that the two-qubit gate is applied only once, in contrast to our first approach (using a transverse field Ising Hamiltonian), or CNOT gates generated by using other types of exchange interaction.\cite{Loss, Schuch, Tanamoto}

Other important two-qubit gates can be achieved by combining the controlled-NOT gate and single-qubit gates. For example, the controlled-$Z$ gate is given by
$U^{ab}_{\rm CZ}=H_{b}U^{ab}_{\rm CNOT}H_{b}$,\cite{Nielsen, Oh, Burkard2} and the swap gate is a combination of three controlled-NOT gate:
$U^{ab}_{\rm SW}=U^{ab}_{\rm CNOT}U^{ba}_{\rm CNOT}U^{ab}_{\rm CNOT}$.\cite{Nielsen}

\section{\label{Sec_V}Discussions and Summary}

In our study, we did not consider qubit decoherence due to the environment. In real quantum dot devices, the main decoherence
sources are the lattice nuclear spins\cite{Cywinski} and the background charge fluctuations.\cite{Hanson, Coish, Hu2}
For a DQD with a finite exchange coupling $J$, charge noise is particularly harmful. In our coupling scheme, however,
we can tune the system to the small $J$ limit in order to minimize this decoherence effect. Furthermore, it may be possible
to remove/suppress decoherence with various dynamical decoupling schemes applied to each qubit (see, e.g.,
Refs. \onlinecite{Barthel2} and \onlinecite{Medford}). In short, it is reasonable to expect that this exchange-based qubit coupling scheme should lead to
better decoherence performance for the $ST_0$ qubits than the capacitive coupling scheme.

The transverse field Ising model has been extensively explored in the areas of adiabatic quantum computing\cite{Murg, Schutzhold, Amin, Wakker} and quantum phase transition.\cite{Sachdev}  Our results here show that the transverse field Ising model can be realized in a tunnel-coupled $ST_{0}$ qubit system, making it a potential platform or quantum simulator for implementing adiabatic quantum computing and investigating quantum phase transitions.

In summary, starting from the recently introduced generalized
Hubbard model,\cite{Yang,Yang2,Wang} we analytically derive a more
complete formula for the exchange coupling between two confined
electrons in a DQD. The tunable exchange coupling depends explicitly
on intradot and interdot Coulomb repulsions, tunnel couplings, and
interdot bias. These two exchange-coupled electron spins can then
be used to encode a singlet-triplet qubit in a strong magnetic
field. We then use the generalized Hubbard model to construct the
spin Hamiltonian for two tunnel-coupled $ST_0$ qubits, and derive a
general expression for the exchange couplings between neighboring
dots in a linear four-dot configuration. This exchange-based
coupling scheme leads to an effective Ising interaction between the
two $ST_{0}$ qubits. The coupling strength is limited by the
magnetic field gradient between the two DQDs, but not limited by the
single-qubit energy splittings, and we estimate that the interqubit
coupling strength can reach the range of 100 MHz. We also explore
how to generate various two-qubit gates with this coupling. In
particular, we discuss how to construct the controlled-NOT gate and
how to generate Bell states.

The results we obtain in our study of a four-quantum-dot system indicate that exchange coupling is a viable
alternative to capacitive coupling for $ST_0$ qubit. It is more
tunable, and its coupling strength is not directly connected to
single-qubit decoherence. With these two favorable features, this exchange-based coupling
scheme for the $ST_0$ qubits deserves closer experimental scrutiny.

{\it Note added}:~As we were finalizing our manuscript, we became
aware of the interesting work by Klinovaja {\it et al.},~\cite{arXiv} which also
dealt with the exchange coupling of $ST_0$ qubits, and included the
effect of spin-orbit coupling as well.

\section*{Acknowledgements}
This work is partially supported by the National Basic Research
Program of China Grant No. 2009CB929302 and the National Natural
Science Foundation of China Grant No. 91121015. X.H. thanks support
by US ARO (W911NF0910393), DARPA QuEST through AFOSR, and NSF PIF
(PHY-1104672).

\appendix

\section{\label{Appendix A}Derivation of the effective interaction Hamiltonian for two electron spins}

\begin{figure}
\includegraphics{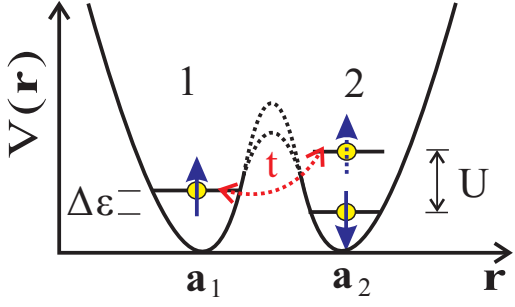}
\caption{\label{Fig_I}(Color online) Double quantum dot model. Two
electrons are confined by a two-well potential $V(\textbf{r})$,
where each well represents a quantum dot and $\textbf{a}_k$ is the
center of the dot $k$. Here only a single electron ground-orbital
energy level is considered for each dot, and $t$ represents the
tunneling coupling between the two dots, which is controllable by
tuning the potential height between the two dots. $U$ is the
intradot Coulomb interaction, and $\Delta\varepsilon$ is the single
electron ground-orbital energy difference between the two dots.}
\end{figure}

In this Appendix we derive the effective spin Hamiltonian for the
low-energy dynamics of two electrons confined in a semiconductor DQD
potential $V(\textbf{r})$, as shown in Fig.~\ref{Fig_I}. The two
electrons interact with each other via the Coulomb interaction and
can be used to achieve the so-called $ST_{0}$ qubit. The effective
mass Hamiltonian for such a two-electron system takes the form
\begin{equation}
H=\sum^{2}_{l=1}\frac{\textbf{p}^{2}_{l}}{2m^{*}_{e}}+\sum^{2}_{l=1}V(\textbf{r}_{l})+\frac{e^{2}}{\epsilon_s |\textbf{r}_{1}-\textbf{r}_{2}|},
\end{equation}
where $m^{*}_{e}$ is the effective mass of the conduction electrons
and $\epsilon_s$ is the static dielectric constant of the material.
Since the total spin is a good quantum number for this Hamiltonian,
the two-electron states can be classified into singlets and
triplets, with symmetric and anti-symmetric orbital wave functions.
The problem can then be solved numerically with the molecular orbit
method (or analytically within the Heitler-London and Hund-Mullikan
approximations).\cite{Burkard, Hu, Scarola} The splitting between
the ground singlet and triplet states is the exchange coupling
between the two confined electrons. With large on-site Coulomb
interaction and single-particle excitation energy, the problem can
also be solved using the Hubbard model, starting with a second
quantized Hamiltonian by defining a field operator
$\Psi(\textbf{r})=\sum^{2}_{k=1,\sigma}c_{k,\sigma}\phi(\textbf{r}-\textbf{a}_{k})\chi_{\sigma}$,
where $\phi(\textbf{r}-\textbf{a}_{k})$ is the single electron
ground orbital wave function in dot $k$ centered at
$\textbf{a}_{k}$,\cite{Burkard,Stepanenko2} and $\chi_{\sigma}$ is
the spin eigenfunction that describes the electron spin degree of
freedom. While in this model higher-energy orbital states are
neglected, it does have the advantage that it can be easily extended
to larger multiple-dot systems. The second quantized Hamiltonian
reads
\begin{eqnarray}
H\!&\!=\!&\!\int\,d\textbf{r}\Psi^{\dagger}(\textbf{r}) \left(\frac{\textbf{p}^{2}}{2m^{*}} + V(\textbf{r}) \right) \Psi(\textbf{r})+\nonumber\\
&&\!\frac{1}{2}\int\,d\textbf{r}d\textbf{r}'
\Psi^{\dagger}(\textbf{r})
\Psi^{\dagger}(\textbf{r}')\frac{e^{2}}{\epsilon_s|\textbf{r}-\textbf{r}'|}
\Psi(\textbf{r}') \Psi(\textbf{r}).~~ \label{SQZH}
\end{eqnarray}
Expressing the Hamiltonian $H$ with the fermionic creation and annihilation
operators for each single-dot single-electron eigenstate, one obtains a generalized Hubbard
model,\cite{Yang,Yang2,Wang}
\begin{eqnarray}
H\!&\!=\!&\!H_{e}+H_{t}+H_{U}+H_{J},\nonumber\\
H_{e}\!&\!=\!&\!\sum^{2}_{k=1,\sigma}\varepsilon_{k}c^{\dagger}_{k\sigma}c_{k\sigma},\nonumber\\
H_{t}\!&\!=\!&\!t\sum_{\sigma}(c^{\dagger}_{1\sigma}c_{2\sigma}+{\rm H.c.}),\nonumber\\
H_{U}\!&\!=\!&\!Un_{1\uparrow}n_{1\downarrow}+Un_{2\uparrow}n_{2\downarrow}+U'(n_{1\uparrow}+n_{1\downarrow})(n_{2\uparrow}+n_{2\downarrow}),
\nonumber\\
H_{J}\!&\!=\!&\!-J_{e}(n_{1\uparrow}n_{2\uparrow}+n_{1\downarrow}n_{2\downarrow})-\big(J_{e}c^{\dagger}_{1\downarrow}c^{\dagger}_{2\uparrow}
c_{2\downarrow}c_{1\uparrow}\nonumber\\
&&\!+J_{p}c^{\dagger}_{2\uparrow}c^{\dagger}_{2\downarrow}c_{1\uparrow}c_{1\downarrow}+\sum_{i\sigma}J_{t}n_{i\sigma}c^{\dagger}_{1\bar{\sigma}}
c_{2\bar{\sigma}}+{\rm H.c.}\big).\label{Eq_Hamiltonian}
\end{eqnarray}
Here $H_{e}$ and $H_{t}$ represent respectively the single
electron energy level of each dot and the tunnel-coupling between
the two dots. Together they make up the single-particle part of the Hamiltonian. The two-body part, i.e., the Coulomb interaction, is
described by $H_{U}$ and $H_{J}$. Previous studies often started
from $H=H_{e}+H_{t}+H_{U}$, which only gives an anti-ferromagnetic
exchange coupling between the two electrons.\cite{Coish} However,
numerical results based on either Heitler-London approximation or
Hund-Mulliken molecular-orbital method have shown that the exchange
coupling can be ferromagnetic in certain parameter
regimes.\cite{Burkard,Hu,Scarola} Here, starting from the general
second quantized Hamiltonian in Eq.~(\ref{Eq_Hamiltonian}), we can obtain
a more precise expression for the exchange coupling between the two electrons.

In the limit of $(U-U')\gg\,t$, there is no charge degree of freedom
for the two electrons; each dot confines one and only one electron
(i.e., double-occupation in each dot is forbidden), and
the two electrons are only allowed in the $(11)$ charge
configuration. Below we derive the effective spin-spin interaction
between the two electrons, i.e., the exchange coupling, by using a
projection operator method. We define a projection operator
\begin{equation}
P=\prod^{2}_{k=1}[n_{k\uparrow}(1-n_{k\downarrow})+n_{k\downarrow}(1-n_{k\uparrow})],
\end{equation}
which projects the Hilbert space of the two electrons to the
subspace $\mathcal{H}$ for $(11)$ charge configuration. Using a standard procedure,\cite{Nagaosa}
we derive an effective spin Hamiltonian that describes the dynamics of $\mathcal{H}$,
\begin{equation}
H_{\rm eff}=PHP-PHQ(QHQ-E)^{-1}QHP,
\end{equation}
where $Q=1-P$. After some algebra, we obtain the following relations:
\begin{eqnarray}
PHP\!&\!=\!&\!\sum^{2}_{k=1}\epsilon_{k}+U'-\frac{1}{2}J_{e}-2J_{e}(\textbf{S}_{1}\cdot\textbf{S}_{2}+\frac{1}{4}),\nonumber\\
PHQ\!&\!=\!&\!PH'_{t}Q,~~~~QHP=QH'_{t}P,
\end{eqnarray}
where
$H'_{t}=(t-J_{t})\sum_{\sigma}(c^{\dagger}_{1\sigma}c_{2\sigma}+{\rm
H.c})$. In other words, under projection operations, the $J_{t}$
term in $H_{J}$ has the same effect as $H_{t}$. In addition, under the lowest order approximation, $QHQ-E$
can be replaced by the difference between the first charge
excitation energy $2\varepsilon_{1}+U$ and the ground state energy
$\varepsilon_{1}+\varepsilon_{2}+U'$, where
$\varepsilon_{2}>\varepsilon_{1}$. Thus, we have
\begin{equation}
QHQ-E\approx\,U-U'-|\Delta\varepsilon|,
\end{equation}
where $\Delta\varepsilon=\varepsilon_{2}-\varepsilon_{1}$. Neglecting the constant terms, we obtain the effective Hamiltonian
\begin{equation}
H_{\rm
eff}=-2J_{e}\textbf{S}_{1}\cdot\textbf{S}_{2}-\frac{(PH'_{t}Q)(QH'_{t}P)}{U-U'-|\Delta\varepsilon|}.
\end{equation}
Using the properties $Q^{2}=Q$, $Q=1-P$, and $PH'_{t}P=0$, it follows that
\begin{equation}
H_{\rm
eff}=-2J_{e}\textbf{S}_{1}\cdot\textbf{S}_{2}-P\frac{H'^{2}_{t}}{U-U'-|\Delta\varepsilon|}P.\label{Hef}
\end{equation}
Now we focus on simplifying the second term in the above equation,
\begin{eqnarray}
PH'^{2}_{t}P\!&\!=\!&\!(t-J_{t})^{2}\sum_{\sigma\sigma'}\big[Pc^{\dagger}_{1\sigma}c_{1\sigma'}PPc_{2\sigma}c^{\dagger}_{2\sigma'}P\nonumber\\
&&\!+Pc^{\dagger}_{2\sigma}c_{2\sigma'}PPc_{1\sigma}c^{\dagger}_{1\sigma'}P\big].
\end{eqnarray}
We can use the following identities:
\begin{eqnarray}
c^{\dagger}_{k\sigma}c_{k\sigma'}\!&\!=\!&\!\frac{1}{2}
\delta_{\sigma\sigma'}(n_{k\uparrow}+n_{k\downarrow})
+\textbf{S}_{k}\cdot\vec{\sigma}_{\sigma'\sigma},\nonumber\\
c_{k\sigma}c^{\dagger}_{k\sigma'}\!&\!=\!&\!\delta_{\sigma\sigma'}
\big[1-\frac{1}{2}(n_{k\uparrow}+n_{k\downarrow})\big]
-\textbf{S}_{k}\cdot\vec{\sigma}_{\sigma\sigma'},~~~
\end{eqnarray}
where the spin operator
$\textbf{S}_{k}=\frac{1}{2}\sum_{\sigma,\sigma'}c^{\dagger}_{k\sigma}\boldsymbol{\sigma}_{\sigma\sigma'}c_{k\sigma'}$,
with $\boldsymbol{\sigma}$ being the Pauli operator, and
$\boldsymbol{\sigma}_{\sigma\sigma'}$ its matrix elements.  The number operators satisfy
$P(\sum_{\sigma}n_{k\sigma})P=1$. After some algebra, we obtain
\begin{eqnarray}
PH'^{2}_{t}P\!&\!=\!&\!\sum_{\sigma\sigma'}2(t-J_{t})^{2}(\frac{1}{2}\delta_{\sigma'\sigma}+\textbf{S}_{1}\cdot\vec{\sigma}_{\sigma'\sigma})\nonumber\\
&&\!\times(\frac{1}{2}\delta_{\sigma\sigma'}-\textbf{S}_{2}\cdot\vec{\sigma}_{\sigma\sigma'})\nonumber\\
\!&\!=\!&\!4(t-J_{t})^{2}(\frac{1}{4}-\textbf{S}_{1}\cdot\textbf{S}_{2}).
\end{eqnarray}
Substituting this equation into Eq.~(\ref{Hef}), and neglecting the
constant terms, we finally obtain the effective interaction
Hamiltonian for the two electron spins
\begin{equation}
H_{\rm eff}=J\textbf{S}_{1}\cdot\textbf{S}_{2},
\end{equation}
where \begin{equation}
J=\frac{4(t-J_{t})^{2}}{U-U'-|\Delta\varepsilon|}-2J_{e}
\end{equation}
is the exchange splitting between the two electrons. These are the
results given in Eqs.~(\ref{EQ_exchange}) and (\ref{eq:exchange
constant}).

\section{\label{Appendix B}Derivation of the effective interaction Hamiltonian for four electron spins}

As discussed in the main text, the coupled qubit system that we
consider is a linearly coupled quadruple quantum dot (see
Fig.~\ref{Fig_II}), with dots 1 and 2 encoding the first qubit, and
dots 3 and 4 encoding the second.  In addition to the coupling
between dots 1 and 2, and that between 3 and 4, we further
allow tunnel coupling between dots 2 and 3, so the two $ST_0$
qubits are tunnel coupled.

Similar to the DQD case discussed in Appendix~\ref{Appendix A}, we define a field
operator $\Psi(\textbf{r}) = \sum^{4}_{k=1,\sigma} c_{k,\sigma}
\phi(\textbf{r} - \textbf{a}_{k}) \chi_{\sigma}$ by using the ground
orbital state $\phi(\textbf{r}-\textbf{a}_{k})$ in each dot. The second quantized Hamiltonian of this quadruple dot
system is then
\begin{eqnarray}
H\!&\!=\!&\!\sum^{4}_{k=1,\sigma}\varepsilon_{k}c^{\dagger}_{k\sigma}c_{k\sigma}
+\sum^{3}_{k=1,\sigma}t_{k,k+1}(c^{\dagger}_{k\sigma}c_{k+1,\sigma}+{\rm H.c.})\nonumber\\
&&\!+U\sum^{4}_{k=1}n_{k\uparrow}n_{k\downarrow}+\sum^{3}_{k=1}H^{k,k+1}_{J}\nonumber\\
&&\!+\sum^{3}_{k=1}U_{k,k+1}(n_{k\uparrow}+n_{k\downarrow})(n_{k+1,\uparrow}+n_{k+1,\downarrow}).
\end{eqnarray}
Here $\varepsilon_{k}$ again describes the single electron ground-orbital
energy in dot $k$, $t_{k,k+1}$ is the tunneling coupling between
two neighboring dots $k$ and $k+1$, $U$ is the intradot Coulomb repulsion,
and both $H^{k,k+1}_{J}$, which is defined in
Eq.~(\ref{Eq_Hamiltonian}), and $U_{k,k+1}$ describe the Coulomb
interactions between two nearest-neighbor dots $k$ and $k+1$. For simplicity, we
consider the symmetric case (between the two $ST_0$ qubits) with the parameters
$U_{12}=U_{34}=U'$, $U_{23}=U''$,
$J^{12}_{e,p,t}=J^{34}_{e,p,t}=J_{e,p,t}$, and
$J^{23}_{e,p,t}=J'_{e,p,t}$. Under the strong Coulomb interaction
conditions, with $(U-U')\gg\,t_{12},t_{34}$ and
$(U-U'')\gg\,t_{23}$, each dot confines only one electron
$\sum_{\sigma}n_{k\sigma}=1$.  Other charge configurations are
not allowed, and these four electrons have only the spin degrees of freedom.

We now apply the same projecting procedure as for a DQD in
Appendix~\ref{Appendix A}, in order to obtain the effective spin Hamiltonian for the
four electrons.  The projection operator is defined as
\begin{equation}
P=\prod^{4}_{k=1}[n_{k\uparrow}(1-n_{k\downarrow})+n_{k\downarrow}(1-n_{k\uparrow})].
\end{equation}
After applying this operator, the Hilbert space for the four electrons is reduced to one involving only the
$(1111)$ charge configuration. Following the same procedure as in
Appendix~\ref{Appendix A}, we obtain the effective interaction
Hamiltonian for the four electron spins
\begin{equation}
H_{\rm
eff}=J_{12}\textbf{S}_{1}\cdot\textbf{S}_{2}+J_{23}\textbf{S}_{2}\cdot\textbf{S}_{3}
+J_{34}\textbf{S}_{3}\cdot\textbf{S}_{4},
\end{equation}
where
\begin{equation}
J_{k,k+1}=\frac{4(t_{k,k+1}-J^{k,k+1}_{t})^{2}}{U-U'+U''-|\Delta\varepsilon_{k,k+1}|}-2J^{k,k+1}_{e}.\label{Eq_exchange_Hamiltonian2}
\end{equation}
These are the results given in Eq.~(\ref{Hamiltonian-four-spins}) and (\ref{Eq_exchange_Hamiltonian2-main}).
The parameters $t_{k,k+1}$, $J^{k,k+1}_{t}$,
$J^{k,k+1}_{e}$, and $\Delta\varepsilon_{k,k+1}$ characterize the
tunneling between nearest-neighbor dots, occupation-modulated
tunneling, spin exchange, and single electron ground-orbital energy
difference, respectively. We emphasize here that all the exchange
splittings $J_{k,k+1}$ are controllable. On the other hand, once the quantum
dot device is designed, the Coulomb-interaction parameters $U$,
$U'$, $U''$, $J_{e}$, and $J_{t}$ are not easily tunable. The directly tunable
parameters $t_{k,k+1}$ and $\Delta\varepsilon_{k,k+1}$ are
controlled by the external gate voltages (see Figs. \ref{Fig_II} and
\ref{Fig_I}).


\begin{thebibliography}{}
\bibitem{Shor}P.W. Shor, \emph{Proceedings of the 35th Annual Symposium on Foundations of Computer Science} (IEEE Computer Soc. Press, Los Alamitos, CA, 1994), p. 124.

\bibitem{Ladd}T.D. Ladd, F. Jelezko, R. Laflamme, Y. Nakamura, C. Monroe, and J.L. O'Brien, Nature (London) {\bf 464}, 45 (2010).

\bibitem{Hanson}R. Hanson, L.P. Kouwenhoven, J.R. Petta, S. Tarucha, and L.M.K. Vandersypen, Rev. Mod. Phys. {\bf 79}, 1217 (2007).

\bibitem{Laird}E.A. Laird, J.R. Petta, A.C. Johnson, C.M. Marcus, A. Yacoby, M.P. Hanson, and A.C. Gossard, Phys. Rev. Lett. {\bf 97}, 056801 (2006).

\bibitem{Morton}J.J.L. Morton, D.R. McCamey, M.A. Eriksson, and S.A. Lyon, Nature (London) {\bf 479}, 345 (2011).

\bibitem{Buluta}I. Buluta, S. Ashhab, and F. Nori, Rep. Prog. Phys. {\bf 74}, 104401 (2011).

\bibitem{Slichter}C.P. Slichter, \emph{Principles of Magnetic Resonance} (Springer-Verlag, Berlin, 1980).

\bibitem{Loss}D. Loss and D.P. DiVincenzo, Phys. Rev. A {\bf 57}, 120 (1998).

\bibitem{Johnson}A.C. Johnson, J.R. Petta, J.M. Taylor, A. Yacoby, M.D. Lukin, C.M. Marcus, M.P. Hanson, and A.C. Gossard, Nature (London) {\bf 435}, 925 (2005).

\bibitem{Petta}J.R. Petta, A.C. Johnson, J.M. Taylor, E.A. Laird, A. Yacoby, M.D. Lukin, C.M. Marcus, M.P. Hanson, and A.C. Gossard, Science {\bf 309}, 2180 (2005).

\bibitem{Koppens}F.H.L. Koppens, J.A. Folk, J.M. Elzerman, R. Hanson, L.H. Willems van Beveren, I.T. Vink, H.P. Tranitz, W. Wegscheider, L.P. Kouwenhoven, L.M.K. Vandersypen, Science {\bf 309}, 1346 (2005).

\bibitem{Tarucha}M. Pioro-Ladriere, T. Obata, Y. Tokura, Y.-S. Shin, T. Kubo, K. Yoshida, T. Taniyama, and S. Tarucha, Nat. Phys. {\bf 4}, 776 (2008).

\bibitem{Foletti}S. Foletti, H. Bluhm, D. Mahalu, V. Umansky, and A. Yacoby, Nat. Phys. {\bf 5} 903 (2009).

\bibitem{Maune}B.M. Maune, M.G. Borselli, B. Huang, T.D. Ladd, P.W. Deelman, K.S. Holabird, A.A. Kiselev, I. Alvarado-Rodriguez, R.S. Ross, A.E. Schmitz, M. Sokolich, C.A. Watson, M.F. Gyure, and A.T. Hunter, Nature (London) {\bf 481}, 344 (2012).

\bibitem{Sachrajda}L. Gaudreau, G. Granger, A. Kam, G.C. Aers, S.A. Studenikin, P. Zawadzki, M. Pioro-Ladriere, Z.R. Wasilewski, and A.S. Sachrajda, Nat. Phys. {\bf 8}, 54 (2011).

\bibitem{Shulman}M.D. Shulman, O.E. Dial, S.P. Harvey, H. Bluhm, V. Umansky, and A. Yacoby, Science {\bf 336}, 202 (2012).

\bibitem{Bluhm}H. Bluhm, S. Foletti, I. Neder, M. Rudner, D. Mahalu, V. Umansky, and A. Yacoby, Nat. Phys. {\bf 7}, 109 (2011).

\bibitem{Lyon}A.M. Tyryshkin, S. Tojo, J.J.L. Morton, H. Riemann, N.V. Abrosimov, P. Becker, H.-J. Pohl, T. Schenkel, M.L.W. Thewalt, K.M. Itoh, and S.A. Lyon, Nat. Mater. {\bf 11}, 143 (2011).

\bibitem{Barthel1}C. Barthel, D.J. Reilly, C.M. Marcus, M.P. Hanson, and A.C. Gossard, Phys. Rev. Lett. {\bf 103}, 160503 (2009).

\bibitem{Barthel2}C. Barthel, J. Medford, C.M. Marcus, M.P. Hanson, and A.C. Gossard, Phys. Rev. Lett. {\bf 105}, 266808 (2010).

\bibitem{Levy}J. Levy, Phys. Rev. Lett. {\bf 89}, 147902 (2002).

\bibitem{Taylor}J.M. Taylor, J.R. Petta, A.C. Johnson, A. Yacoby, C.M. Marcus, and M.D. Lukin, Phys. Rev. B {\bf 76}, 035315 (2007).

\bibitem{Taylor2}J.M. Taylor, H.-A. Engel, W. Dur, A. Yacoby, C.M. Marcus, P. Zoller, and M.D. Lukin, Nat. Phys. {\bf 1}, 177 (2005).

\bibitem{Stepanenko}D. Stepanenko and G. Burkard, Phys. Rev. B {\bf 75}, 085324 (2007).

\bibitem{Ramon1}G. Ramon and X. Hu, Phys. Rev. B {\bf 81}, 045304 (2010).

\bibitem{Ramon2}G. Ramon, Phys. Rev. B {\bf 84}, 155329 (2011).

\bibitem{ENielsen}E. Nielsen, R.P. Muller, and M.S. Carroll, Phys. Rev. B {\bf 85}, 035319 (2012).

\bibitem{Trifunovic}L. Trifunovic, O. Dial, M. Trif, J.R. Wootton, R. Abebe, A. Yacoby, and D. Loss, Phys. Rev. X {\bf 2}, 011006 (2012).

\bibitem{Burkard}G. Burkard, D. Loss, and D.P. DiVincenzo, Phys. Rev. B {\bf 59}, 2070 (1999).

\bibitem{Hu}X. Hu and S. Das Sarma, Phys. Rev. A {\bf 61}, 062301 (2000).

\bibitem{Scarola}V.W. Scarola and S. Das Sarma, Phys. Rev. A {\bf 71}, 032340 (2005).

\bibitem{Coish}W.A. Coish and D. Loss, Phys. Rev. B {\bf 72}, 125337 (2005).

\bibitem{Yang}S. Yang, X. Wang, and S. Das Sarma, Phys. Rev. B {\bf 83}, 161301(R) (2011).

\bibitem{Yang2}S. Yang and S. Das Sarma, Phys. Rev. B {\bf 84}, 121306(R) (2011).

\bibitem{Wang}X. Wang, S. Yang, and S. Das Sarma, Phys. Rev. B {\bf 84}, 115301 (2011).

\bibitem{Obata}T. Obata, M. Pioro-Ladriere, Y. Tokura, Y.-S. Shin, T. Kubo, K. Yoshida, T. Taniyama, and S. Tarucha, Phys. Rev. B {\bf 81}, 085317 (2010).

\bibitem{Brunner}R. Brunner, Y.-S. Shin, T. Obata, M. Pioro-Ladriere, T. Kubo, K. Yoshida, T. Taniyama, Y. Tokura, and S. Tarucha, Phys. Rev. Lett. {\bf 107}, 146801 (2011).

\bibitem{Hanson2}R. Hanson, G. Burkard, Phys. Rev. Lett. {\bf 98}, 050502 (2007).

\bibitem{Assali}L.V.C. Assali, H.M. Petrilli, R.B. Capaz, B. Koiller, X. Hu, and S. Das Sarma, Phys. Rev. B {\bf 83}, 165301 (2011).

\bibitem{Nielsen}M.A. Nielsen and I.L. Chuang, \emph{Quantum Computations and Quantum Information} (Cambridge University Press, Cambridge, 2002).

\bibitem{Makhlin}Y. Makhlin, Quantum Inf. Processing {\bf 1}, 243 (2002).

\bibitem{Oh}S. Oh, Phys. Rev. B {\bf 65}, 144526 (2002).

\bibitem{Hahn}E.L. Hahn, Phys. Rev. {\bf 80}, 580 (1950).

\bibitem{Schuch}N. Schuch and J. Siewert, Phys. Rev. A {\bf 67}, 032301 (2003).

\bibitem{Tanamoto}T. Tanamoto, Y.X. Liu, X. Hu, and F. Nori, Phys. Rev. Lett. {\bf 102}, 100501 (2009).

\bibitem{Burkard2} G. Burkard, D. Loss, D.P. DiVincenzo, and J.A. Smolin, Phys. Rev. B {\bf 60} 11404 (1999).

\bibitem{Cywinski}L. Cywinski, W.M. Witzel, and S. Das Sarma, Phys. Rev. B {\bf 79}, 245314 (2009).

\bibitem{Hu2}X. Hu and S. Das Sarma, Phys. Rev. Lett. {\bf 96}, 100501 (2006).

\bibitem{Medford}J. Medford, L. Cywinski, C. Barthel, C.M. Marcus, M.P. Hanson, and A.C. Gossard, Phys. Rev. Lett. {\bf 108}, 086802 (2012).

\bibitem{Murg}V. Murg and J.I. Cirac, Phys. Rev. A {\bf 69}, 042320 (2004).

\bibitem{Schutzhold}R. Schutzhold and G. Schaller, Phys. Rev. A {\bf 74}, 060304(R) (2006).

\bibitem{Amin}M.H.S. Amin and V. Choi, Phys. Rev. A {\bf 80}, 062326 (2009).

\bibitem{Wakker}G.M.M. Wakker, R. Ockhorst, and M. Blaauboer, Phys. Rev. A {\bf 85}, 022337 (2012).

\bibitem{Sachdev}S. Sachdev, \emph{Quantum Phase Transitions} (Cambridge University Press, Cambridge, 1999).

\bibitem{arXiv}J. Klinovaja, D. Stepanenko, B.I. Halperin, Daniel Loss, Phys. Rev. B {\bf 86}, 085423 (2012).

\bibitem{Stepanenko2}D. Stepanenko, M. Rudner, B.I. Halperin, and D. Loss, Phys. Rev. B {\bf 85}, 075416 (2012).

\bibitem{Nagaosa} N. Nagaosa, \emph{Quantum Field Theory in Strongly Correlated Electronic Systems} (Springer-Verlag, Berlin, 1999), p. 79.

\end{thebibliography}
\end{document}